\documentclass[a4paper,12pt,reqno]{amsart}

\usepackage[centertags]{amsmath}

\usepackage{mathrsfs}
\usepackage{amsfonts}
\usepackage{amssymb}
\usepackage{amsthm}
\usepackage{newlfont}
\usepackage{stmaryrd}

\theoremstyle{plain}
  \newtheorem{theorem}{Theorem}[section]
  
  \newtheorem{proposition}[theorem]{Proposition}
  \newtheorem{lemma}[theorem]{Lemma}
\theoremstyle{definition}

\theoremstyle{remark}
  \newtheorem{remark}[theorem]{Remark}

\numberwithin{equation}{section}

\let\al=\alpha  \let\de=\delta \let\ep=\epsilon
   
 \let\la=\lambda \let\om=\omega 
\let\si=\sigma

  \let\La=\Lambda \let\Om=\Omega

\newcommand{\caF}{{\mathcal F}}

\newcommand{\caH}{{\mathcal H}}
\newcommand{\caI}{{\mathcal I}}

\newcommand{\caZ}{{\mathcal Z}}

\newcommand{\bbC}{{\mathbb C}}

\newcommand{\bbN}{{\mathbb N}}

\newcommand{\bbR}{{\mathbb R}}

\newcommand{\opunit}{\text{1}\kern-0.22em\text{l}}

\newcommand{\rel}{\,|\,}

\newcommand{\id}{\textrm{d}}

\newcommand{\tomean}{\stackrel{1}{\to}}
\newcommand{\ra}{\rightarrow}

\newcommand{\tr}{\mathrm{tr}}
\newcommand{\Tr}{\mathrm{Tr}}

\newcommand{\can}{\mathrm{can}}
\newcommand{\mc}{\mathrm{mc}}
\newcommand{\tomc}{\stackrel{\mathrm{mc}}{\to}}

\newcommand{\beq}{ \begin{equation} }
\newcommand{\eeq}{ \end{equation} }
\newcommand{\bet}{ \begin{theorem} }
\newcommand{\eet}{ \end{theorem} }

\newcounter{smallarabics}
\newenvironment{arabicenumerate}
{\begin{list}{{\normalfont\textrm{\arabic{smallarabics})}}}
  {\usecounter{smallarabics}\setlength{\itemindent}{0cm}
  \setlength{\leftmargin}{5ex}\setlength{\labelwidth}{4ex}
  \setlength{\topsep}{0.75\parsep}\setlength{\partopsep}{0ex}
   \setlength{\itemsep}{0ex}}}
{\end{list}}

\newcounter{smallroman}

\newcommand{\ben}{\begin{arabicenumerate}}
\newcommand{\een}{\end{arabicenumerate}}

\begin{document}
\begin{center}
\noindent{\large \bf  Quantum Macrostates, Equivalence of Ensembles\\
and an $H-$Theorem} \\

\vspace{15pt}

{\bf Wojciech De Roeck}\footnote{Aspirant FWO, U.Antwerpen}, {\bf Christian Maes}\footnote{email:
{\tt
christian.maes@fys.kuleuven.be}} \\
Instituut voor Theoretische Fysica, K.U.Leuven\\\vspace{10pt} {\bf Karel
Neto\v{c}n\'{y}\footnote{email: {\tt netocny@fzu.cz}}}\\
Institute of Physics AS CR, Prague
\end{center}
\vspace{15pt} Dedicated to Andr\'e Verbeure on the occasion of his
65th birthday.

\vspace{20pt} \footnotesize \noindent {\bf Abstract: } Before the
thermodynamic limit, macroscopic averages need not commute for a
quantum system. As a consequence, aspects of macroscopic
fluctuations or of constrained equilibrium require a careful
analysis, when dealing with several observables. We propose an
implementation of ideas that go back to John von Neumann's writing
about the macroscopic measurement.  We apply our scheme to the
relation between macroscopic autonomy and an $H-$theorem, and to
the problem of equivalence of ensembles. In particular, we show
how the latter is related to the asymptotic equipartition theorem.
The main point of departure is an expression of a law of large
numbers for a sequence of states that start to concentrate, as the
size of the system gets larger, on the macroscopic values for the
different macroscopic observables. Deviations from that law are
governed by the entropy.

\vspace{5pt}
\footnotesize \noindent {\bf KEY WORDS:}
quantum macrostate, autonomous equations, $H-$theorem, equivalence of ensembles
\vspace{20pt} \normalsize

\section{Introduction}
``It is a fundamental fact with macroscopic measurements that everything which is measurable at
all, is also simultaneously measurable, i.e. that all questions which can be answered separately
can also be answered simultaneously.''  That statement by von Neumann enters his introduction to
the macroscopic measurement \cite{vN}. He then continues to discuss in more detail how that view
could possibly be reconciled with the non-simultaneous measurability of quantum mechanical
quantities. The mainly qualitative suggestion by von Neumann is to consider, for a set of
noncommuting operators $A,B,...$ a corresponding set of {\it mutually commuting} operators
$A',B',\ldots$ which are each, in a sense, good approximations,
$A'\approx A, B'\approx B,\ldots$.
The whole question is: {\it in exactly what sense?} Especially in statistical mechanics, one is
interested in fluctuations of macroscopic quantities or in the restriction of certain ensembles by
further macroscopic constraints which only make sense for finite systems. In these cases, general
constructions of a common subspace of observables become very relevant. Interestingly, at the end
of his discussion on the macroscopic measurement,
\cite{vN}, von Neumann turns to the quantum $H-$theorem and to the
relation between entropy and macroscopic measurement. He refers to the then recent work of Pauli,
\cite{Pauli, Tolman}, who by using ``disorder assumptions'' or what we could call today, a
classical Markov approximation, obtained a general argument for the $H-$theorem.

In the present paper, we are dealing exactly with the problems above and as discussed in Chapter
V.4 of \cite{vN}.  While it is indeed true that averages of the form $A = (a_1 +\ldots + a_N)/N,
B= (b_1 + \ldots + b_N)/N$, for which all commutators $[a_i,b_j] =0$ for $i \neq j$, have their
commutator $[A,B] = O(1/N)$ going to zero (in the appropriate norm, corresponding to $[a_i,b_i] =
O(1)$) as $N\uparrow +\infty$, it is not true in general that
\[
 \lim_{N \ra +\infty} \frac 1{N} \log \mbox{Tr} [e^{NA}\, e^{NB}] \stackrel{?}{=}
\lim_{N \ra +\infty} \frac 1{N} \log \mbox{Tr} [e^{NA +NB}]
\]
These generating functions are obviously important in fluctuation theory, such as in the problem
of large deviations for quantum systems, \cite{NR}.  It is still very much an open question to
discuss the joint large deviations of quantum observables, or even to extend the Laplace-Varadhan
formula to applications in quantum spin systems. The situation is better for questions about
normal fluctuations and the central limit theorem, for which the so-called fluctuation algebra
provides a nice framework, see e.g. \cite{av}. There the pioneering work of Andr\'e Verbeure will
continue to inspire coming generations who are challenged by the features of non-commutativity
in quantum mechanics.\\
These issues are also important for the question of convergence to equilibrium.  For example, one
would like to specify or to condition on various macroscopic values when starting off the system.
Under these constrained equilibria not only the initial energy but also e.g. the initial
magnetization or particle density etc. are known, and simultaneously installed.  As with the large
deviation question above, we enter here again in the question of equivalence of ensembles but we
are touching also a variety of problems that deal with nonequilibrium aspects.  The very
definition of configurational entropy as related to the size of the macroscopic subspace, has to
be rethought when the macroscopic variables get their representation as noncommuting operators.
One could again argue that all these problems vanish in the macroscopic limit, but the question
(indeed) arises before the limit, for very large but finite $N$ where one can still speak about
finite dimensional subspaces or use arguments like the
Liouville-von Neumann theorem.\\

In the following, there are three sections.  In Section 2 we write
about quantum macrostates and about how to define the macroscopic
entropy associated to values of several noncommuting observables.
As in the classical case, there is the Gibbs equilibrium entropy.
The statistical interpretation, going back to Boltzmann for
classical physics, is however not immediately clear in a quantum
context. We will define various quantum $H-$functions. Secondly,
in Section 3, we turn to the equivalence of ensembles. The main
result there is to give a counting interpretation to the
thermodynamic equilibrium entropy. In that light we discuss
quantum aspects of large deviation theory. Finally, in Section 4,
we study the relation between macroscopic autonomy and the second
law, as done before in \cite{dmn} for classical dynamical systems.
We prove that if the macroscopic observables give rise to a first
order autonomous equation, then the $H-$function, defined on the
macroscopic values, is monotone. That is further illustrated using
a quantum version of the Kac ring model.

\section{Quantum macrostates and entropy}\label{sec: macrostates}

Having in mind a macroscopically large closed quantum dynamical system, we consider a sequence
$\mathscr{H} = (\mathscr{H}^N)_{N\uparrow +\infty}$ of finite-dimensional Hilbert spaces with the index $N$
labeling different finitely extended approximations, and playing the role of the volume or the
particle number, for instance. On each space
$\mathscr{H}^N$ we have the standard trace $\Tr^N$.
Macrostates are usually identified with subspaces of the Hilbert spaces or, equivalently, with the
projections on these subspaces. For any collection $(X^N_k)_{k=1}^n$ of mutually commuting
self-adjoint operators there is a projection-valued measure
$(Q^N)$ on $\bbR^n$ such that for any
function $F \in C(\bbR^n)$,
\[
  F(X^N_{1},\ldots,X^N_{n}) = \int_{\bbR^n} Q^N(\id z)\,F(z)
\]
A macrostate corresponding to the respective values
$x=(x_1,x_2,\ldots,x_n)$ is then represented by the projection
\[
  Q^{N,\de}(x) = \int_{\vartimes_k (x_k - \de, x_k + \de)} Q^N(\id z)
\]
for small enough $\delta > 0$. Furthermore, the Boltzmann
$H-$function, in the classical case counting the cardinality of
macrostates, is there defined as
\[
  H^{N,\de}(x) = \frac{1}{N} \log \Tr^N[Q^{N,\de}(x)]
\]
with possible further limits $N\uparrow +\infty$, $\delta\downarrow 0$. However, a less trivial
problem that we want to address here, emerges if the observables $(X^N_k)$ chosen to describe the
system on a macroscopic scale do not mutually commute.\\

Consider a family of sequences of self-adjoint observables
$(X^N_k)_{N\uparrow +\infty, k \in K}$ where $K$ is some index
set, and let each sequence be uniformly bounded, $\sup_N \|X^N_k\| < +\infty$, $k \in K$.  We call
these observables macroscopic, having in mind mainly averages of local observables but that will
not always be used explicitly in what follows; it will however serve to make the assumptions
plausible.

In what follows, we define concentrating states as sequences of states for which the observables
$X^N_k$ assume sharp values. Those concentrating states will be labeled by possible `outcomes' of
the observables $X^N_k$; for these values we write $x = (x_k)_{k\in K}$ where each $x_k \in \bbR$.

\subsection{Microcanonical set-up}\label{sec21}

\subsubsection{Concentrating sequences}\label{cs}
A sequence $(P^N)_{N\uparrow +\infty}$ of projections is called concentrating at $x$ whenever
\begin{equation}\label{eq: concentration}
  \lim_{N\uparrow +\infty} \tr^N(F(X^N_k) \rel P^N) = F(x_k)
\end{equation}
for all $F \in C(\bbR)$ and $k\in K$; we have used the notation
\begin{equation}
  \tr^N(\cdot \rel P^N) = \frac{\Tr^N(P^N\cdot P^N)}{\Tr^N(P^N)}
  = \frac{\Tr^N(P^N\cdot)}{\Tr^N(P^N)}
\end{equation}
for the normalized trace state on $P^N \mathscr{H}^N$. To indicate that a sequence of projections
is concentrating at $x$ we use the shorthand $P^N \tomc x$.

\subsubsection{Noncommutative functions}
The previous lines, in formula \eqref{eq: concentration}, consider functions of a single
observable. By properly defining the joint functions of two or more operators that do not mutually
commute, the concentration property extends as follows.\\
Let $\caI_K$ denote the set of all finite sequences from $K$, and consider all maps $G: \caI_K \ra
\bbC $ such that
\begin{equation}
  \sum_{m\geq 0} \sum_{(k_1,\ldots,k_m) \in \caI_K} |G(k_1,\ldots,k_m)| \prod_{i=1}^m r_{k_i} < \infty
\end{equation}
for some fixed $r_k > \sup_N \|X^N_k\|, k \in K$. Slightly abusing the notation, we also write
\begin{equation}
  G(X^N) = \sum_{m \geq 0} \sum_{(k_1,\ldots,k_m) \in \caI_K} G(k_1,\ldots,k_m)\,
  X^N_{k_1} \ldots X^N_{k_m}
\end{equation}
defined as norm-convergent series. We write $\caF$ to denote the algebra of all these maps $G$,
defining non-commutative ``analytic'' functions on the multidisc with radii $(r_k), k\in K$.

\begin{proposition}\label{prop: noncommutative functions}
Assume that $P^N \tomc x$.  Then, for all $G \in \caF$,
\begin{equation}\label{eq: concentration for non-commutative functions}
  \lim_{N\uparrow +\infty} \tr^N[G(X^N) \rel P^N] = G(x)
\end{equation}
\end{proposition}
\begin{remark}
In particular, the limit expectations on the left-hand side of
\eqref{eq: concentration for non-commutative functions} coincide
for all classically equivalent non-commutative functions. As example, for any complex parameters
$\la_k, k\in R$ with $R$ a finite subset of $K$ and for $P^N \tomc
x$,
\begin{equation*}
\begin{split}
  \lim_{N\uparrow +\infty} \tr^N(e^{\sum_{k\in R} \la_k (X^N_k - x_k)} \rel P^N)
  &= \lim_{N\uparrow +\infty} \tr^N(\prod_{k\in R}e^{\la_k (X^N_k - x_k)}\rel P^N)
  = 1
\end{split}
\end{equation*}
no matter in what order the last product is actually performed.
\end{remark}

\begin{proof}[Proof of Proposition~\ref{prop: noncommutative functions}]
For any monomial $G(X^N) = X^N_{k_1} \ldots X^N_{k_m}$, $m \geq
1$, we prove the statement of the proposition by induction, as
follows. Using the shorthands $Y^N = X^N_{k_1} \ldots
X^N_{k_{m-1}}$ and $y = x_{k_1} \ldots x_{k_{m-1}}$, the induction
hypothesis reads $\lim_{N \uparrow +\infty} \rho^N(Y^N \rel P^N) =
y$ and we get
\begin{equation*}
\begin{split}
  |\tr^N&(Y^N X^N_{k_m} - y x_{k_m} \rel P^N)|
\\
  &= |\tr^N(Y^N (X_{k_m}^N - x_{k_m}) \rel P^N) + x_{k_m} \tr^N(Y^N - y \rel P^N)|
\\
  &\leq \|Y^N\|\{\tr^N((X^N_{k_m} - x_{k_m})^2 \rel P^N)\}^{\frac{1}{2}}
  + |x_{k_m}|\,|\tr^N(Y^N - y \rel P^N)| \to 0
\end{split}
\end{equation*}
since $P^N \tomc x$ and $(Y^N)$ are uniformly bounded. That readily extends to all non-commutative
polynomials by linearity, and finally to all uniform limits of the polynomials by a standard
continuity argument.
\end{proof}

\subsubsection{$H-$function}
Only the concentrating sequences of projections on the subspaces of the largest dimension become
candidates for non-commutative variants of macrostates associated with $x = (x_k)_{k \in K}$, and
that maximal dimension yields the (generalization of) Boltzmann's
$H-$function. More precisely, to any macroscopic value $x =
(x_k)_{k \in K}$ we assign
\begin{equation}\label{eq: H-function}
  H^{\mc}(x) = 
\limsup_{P^N \tomc x} \frac{1}{N} \log \Tr^N[P^N]
\end{equation}
where $\limsup_{P^N \tomc x} = \sup_{P^N \tomc x} \limsup_{N
\uparrow +\infty}$ is the maximal limit point over all sequences
of projections concentrating at $x$. By construction, $H^\mc(x)
\in \{-\infty\} \cup [0,+\infty]$  and we write $\Om$ to denote
the set of all $x \in \bbR^K$ for which $H^\mc(x) \geq 0$; these
are all admissible macroscopic configurations. Slightly abusing
the notation, any sequence $P^N \tomc x$, $x \in \Om$ such that
$\limsup_{N} \frac{1}{N} \log\Tr^N[P^N] = H^{\mc}(x)$, will be
called a microcanonical macrostate at $x$.

\subsubsection{Example}
Take a spin system of $N$ spin-$1/2$ particles for which the magnetization in the
$\alpha-$direction, $\alpha=1,2,3$, is given by
\begin{equation}
X^N_{\alpha}=  \frac{1}{N} \sum_{i=1}^N \sigma^{\alpha}_i
\end{equation}
in terms of (copies of) the Pauli matrices $\sigma^\alpha$. \\
Let $\delta_N$ be a sequence of positive real numbers such that
$\de_N \downarrow 0$ as $N\uparrow +\infty$. For
$\vec{m}=(m_1,m_2,m_3) \in [-1,1]^3$, let $\vec{e}\parallel
\vec{m}$ be a unit vector for which $\vec m = m\,\vec{e}$ with $m
\geq 0$. Consider $Y^N(\vec m) = \sum_{\alpha=1}^3 m_{\al}
X_\alpha^N$ and its spectral projection $Q^{N}(\vec m)$  on
$[m-\de_N,m+\de_N]$. One easily checks that if $N^{1/2}\de_N
\uparrow +\infty$, then $(Q^{N}(\vec m))_N$ is a microcanonical
macrostate at $\vec{m}$, and
\[
  H^\mc(\vec m) =
  \begin{cases}
    -\frac{1-m}{2}\log\frac{1-m}{2} - \frac{1+m}{2}\log\frac{1+m}{2}
    & \text{for } m \leq 1
  \\
    -\infty
    & \text{otherwise}
  \end{cases}
\]

\subsection{Canonical set-up}\label{cans} The concept of macrostates
as above and associated with projections on certain subspaces on which the selected macroscopic
observables take sharp values is physically natural and restores the interpretation of ``counting
microstates''.  Yet, sometimes it is not very suitable for computations. Instead, at least when
modeling thermal equilibrium, one usually prefers canonical or grand-canonical ensembles, and one
relies on certain equivalence of all these ensembles.

\subsubsection{Concentrating states}
For building the ensembles of quantum statistical mechanics, one does not immediately encounter
the problem of noncommutativity. One requires a certain value for a number of macroscopic
observables and one constructs the density matrix that maximizes
the von Neumann entropy.\\
We write $\om^N \tomean x$ for a sequence of states $(\omega^N$) on $\mathscr{H}^N$ whenever
$\lim_{N \uparrow +\infty} \om^N(X^N_k) = x_k$ (convergence in
mean).\\
That construction and that of the concentrating sequences of projections of subsection \ref{cs}
still has other variants.  We say that a sequence of states $(\om^N)$ is concentrating at $x$ and
we write $\om^N \to x$, when
\begin{equation}
  \lim_{N\uparrow +\infty} \om^N(G(X^N)) = G(x)
\end{equation}
for all $G \in \caF$. The considerations of Proposition \ref{prop: noncommutative functions} apply
also here and one can equivalently replace the set of all noncommutative analytic functions with
functions of a single variable.

\subsubsection{Gibbs-von Neumann entropy}

The counting entropy of Boltzmann extends to general states as the von Neumann entropy which is
the quantum variant of the Gibbs formula, both being related to the relative entropy defined with
respect to a trace reference state. Analogously to \eqref{eq: H-function}, we define
\begin{equation}\label{rela}
  H^\can(x) = \limsup_{\om^N \to x} \frac{1}{N} \caH(\om^N)
\end{equation}
where $\caH(\om^N) \geq 0$ is, upon identifying the density matrix $\si^N$ for which
$\om^N(\cdot) = \Tr^N(\si^N \cdot)$,
\begin{equation}\label{eq: G.-von N.}
  \caH(\om^N) = -\Tr [\si^N \log \si^N]
\end{equation}

Secondly, we consider
\begin{equation}\label{h1}
  H^\can_1(x) = \limsup_{\om^N \tomean x} \frac{1}{N} \caH(\om^N)
\end{equation}
 Obviously, $H^\can_1$ is
the analogue of the canonical entropy in thermostatics and the easiest to compute, see also under
subsection \ref{cmex}. To emphasize that, we call any sequence of states $(\om^N)$, $\om^N
\tomean x$ such that
$\limsup_N \frac{1}{N} \caH(\om^N) = H^\can_1(x)$
a canonical macrostate at $x$.

Another generalization of the $H-$function is obtained when replacing the trace state
(corresponding to the counting) with a more general reference state $\rho = (\rho^N)_N$. In that
case we consider the $H-$function as derived from the relative entropy, and differing from the
convention used above by the sign and an additive constant:
\begin{equation}\label{addi}
  H^\can_1(x \rel \rho) = \liminf_{\om^N \tomean x} \frac{1}{N} \caH(\om^N \rel \rho^N)
\end{equation}
Here, defining $\si^N$ and $\si^N_0$ as the density matrices such that
$\om^N(\cdot) = \Tr[\si^N \cdot]$ and $\rho^N(\cdot) = \Tr[\si_0^N \cdot]$,
\begin{equation}
  \caH(\om^N \rel \rho^N) = \Tr [\si^N(\log \si^N - \log\si_0^N)]
\end{equation}
Remark that this last generalization enables to cross the border between closed and open
thermodynamic systems. Here, the state
$(\rho^N)$ can be chosen as a nontrivial stationary state for an
open system, and the above defined $H-$function $H^\can_1(x \rel
\rho)$ may loose natural counting and thermodynamic
interpretations. Nevertheless, its monotonicity properties under dynamics satisfying suitable
conditions justify this generalization, see Section~\ref{sec: H-theorems}.

\subsubsection{Canonical macrostates}\label{cmex} The advantage of
the canonical formulation of the variational problem for the
$H-$function as in \eqref{h1}  is that it can often be solved in a
very explicit way. A class of general and well-known examples of
canonical macrostates have the following Gibbsian form.\\
If $\la = (\la_1,\ldots,\la_n)$ are such that the sequence of states $(\om^N_\la)$,
$\om^N_\la(\cdot) = \Tr^N(\si^N_\la \cdot)$ defined by
\begin{equation}\label{eq: explicit}
  \si^N_\la = \frac{1}{\caZ^N_\la}\, e^{N\sum_k \la_k X^N_k} \qquad
  \caZ^N_\la = \Tr^N(e^{N\sum_k \la_k X^N_k})
\end{equation}
satisfies $\lim_{N \uparrow +\infty} \om^N_\la(X^N_k) = x_k$, $k = 1,\ldots,n$, then $(\om^N_\la)$
is a canonical macrostate at $x$, and
\begin{equation}
  H_1^\can(x) = \limsup_N \frac{1}{N} \log\caZ^N_\la - \sum_k \la_k x_k
\end{equation}

\section{Equivalence of ensembles}\label{sec: generalized macrostates}
A basic intuition of statistical mechanics is that adding those
many new concentrating states in the variational problem, as done
in the previous section \ref{cans}, does not actually change the
value of the $H-$function. In the same manner of speaking, one
would like to understand the definitions \eqref{rela} and
\eqref{h1} in counting-terms.  In what sense do these entropies
represent a dimension (the size) of a (microscopic) subspace?\\

Trivially, $H^{\mc} \leq H^\can \leq H^\can_1$, and $H^\can(x) = H^\can_1(x)$ iff some canonical
macrostate  $\om^N \tomean x$ is actually concentrating at $x$, $\om^N \to x$. We give general
conditions under which the full equality can be proven.\\
We have again a sequence of observables $X_k^N$ with spectral measure given by the projections
$Q_k^N(\id z), k\in K$.
\begin{theorem}\label{thm: equivalence}
Assume that for a sequence of density matrices $\sigma^N > 0$, the corresponding $(\om^N)_N$ is a
canonical macrostate  at $x$ and that the  following two conditions are verified:
\begin{enumerate}
\item[i)]
\emph{(Exponential concentration property.)}\\
For every $\de > 0$ and $k\in K$ there are $C_k(\de) > 0$ and
$N_k(\de)$ so that
\begin{equation}\label{excp}
  \int_{x_k - \de}^{x_k + \de} \om^N(Q^N_k(\id z)) \geq 1 - e^{-C_k(\de)N}
\end{equation}
for all $N > N_k(\de)$. \item[ii)]
\emph{(Asymptotic equipartition property.)}\\
For all $\delta>0$,
\begin{equation}\label{eq: AEP}
  \lim_{N\uparrow +\infty} \frac{1}{N} \log\int_{-\de}^{\de} \om^N( \tilde Q^N(dz)) = 0
\end{equation}
where $\tilde Q^N$ denotes the projection operator-valued measure of the operator $\frac{1}{N}
(\log\si^N - \om^N(\log\si^N))$.
\end{enumerate}
Then, $H^{\mc}(x) = H^\can(x) = H^\can_1(x) \geq 0$. \\
\end{theorem}

Theorem \ref{thm: equivalence} evidently expresses that the
microcanonical and the canonical ensembles are equivalent. Results
of that kind are well-known in the literature, see e.g.\
\cite{simon} or \cite{georgii}. An example of a similar type of
reasoning for the quantum case is given in \cite{lima}. Theorem
\ref{thm: equivalence} is however slightly different from these
results in the following aspects,
\begin{enumerate}
\item{When considering the quantum microcanonical ensemble, one
usually starts out with spectral projections $P^N$ associated with
one macroscopic observable. That at least is the approach in
\cite{lima} and it is also sketched at the very beginning of
Section \ref{sec: macrostates}. Our approach is however not
limited to one macroscopic observable. Indeed, remember that the
$(X^N_k)_k$ need not commute (Section \ref{sec21}). }
 \item{Results on equivalence of ensembles, including those contained in
 e.g.~\cite{simon,georgii,lima} are mostly dealing solely with
translation-invariant lattice spin systems. We do not have that
limitation here; instead we have the assumptions \eqref{eq: AEP}
and \eqref{excp}.} \item{Even within the context of
translation-invariant lattice spin systems, the results in
\cite{simon,georgii,lima} do not yield Theorem \ref{thm:
equivalence}. In these references the microcanonical state is
defined as the average of projections $P^N$, translated over all
lattice vectors. That lattice average is translation-invariant by
construction (and hence technically easier to handle), but of
course it is itself not longer a projection and hence it is not a
microcanonical state in the sense of the present paper.}
\end{enumerate}

\subsection*{Remarks on the conditions of Theorem \ref{thm: equivalence}}
Whether one can prove the assumptions of Theorem \ref{thm: equivalence}, depends heavily on the
particular model.

The exponential concentration property \eqref{excp} is not trivial
even for quantum lattice spin systems, and not even in their
one-phase region. Let us mention one criterion under which
\eqref{excp} can be checked, which indicates its deep relation to
the problem of quantum large deviations. Consider the generating
functions
\begin{equation}\label{eq: gen. function}
  \psi_k(t) = \lim_{N\uparrow +\infty} \frac{1}{N} \log \om^N(e^{t N
  X^N_k}),
  \qquad k \in K
\end{equation}
Their existence together with their differentiability at $t=0$
imply by an exponential Chebyshev inequality that $\om^N$
exponentially concentrates at $x = (\psi'_k(0);\,k\in K)$.
However, to our knowledge, the differentiability of $\psi_k(t)$
has only been proven so far for lattice averages over local
observables for quantum spin lattice systems in a
``high-temperature regime'', see \cite{NR}, Theorem 2.15 and
Remark 7.13, where a cluster expansion technique has been used.
The existence of the generating functions \eqref{eq: gen.
function} has also been studied in \cite{Rey-bellet}.

The asymptotic equipartition property \eqref{eq: AEP} is easier. The terminology, originally in
information theory, comes from its immediate
consequence \eqref{eq: equiv-lem} below, where $P^N$
projects on a ``high probability'' region: as in the classical
case, the Gibbs-von Neumann entropy
measures in some sense the size of the space of ``sufficiently probable'' microstates.\\
For \eqref{eq: AEP} it is enough to prove that the state
$\omega^N$ is concentrating for the observable
\begin{equation}
A^N = \frac{1}{N}\log{\sigma^N}
\end{equation}
Explicitly, it is enough to show that for all $F \in C(\bbR)$,
\begin{equation}
\lim_{N \uparrow +\infty} \big[\omega^N(F(A^N))-F(\omega^N(A^N))
\big]=0
\end{equation}
In particular, if $(\om^N)$, $\om^N = \om^N_\la$ is given by formula \eqref{eq: explicit}, a
sufficient condition for the asymptotic equipartition
property to be satisfied is that the pressure
$p(\la)$ defined as
\begin{equation}
  p(\la) = \lim_{N\uparrow +\infty} \frac{1}{N} \log \caZ^N_\la
\end{equation}
exists and is continuously differentiable at $\la = \la(x)$.

Remark that for ergodic states of spin lattice systems, the
asymptotic equipartition as expressed
by~\eqref{eq: AEP} and
\eqref{eq: equiv-lem} follows from the quantum Shannon-McMillan
theorem, see~\cite{deuschel1} and the references therein. An interesting variant of that result
which touches the problem of quantum large deviations, is the quantum Sanov theorem, proven for
i.i.d.\ processes in~\cite{deuschel}. In contrast, our result focuses on the intimate relation of
the asymptotic equipartition property to the problem of equivalence of ensembles in the
noncommutative context, and Theorem~\ref{thm: equivalence} formulates sufficient conditions under
which such an equivalence follows. An advantage of this approach is that it is not restricted to
the framework
of spin lattice models with its underlying quasilocal structure.\\

As $H^{\mc} \leq H^\can \leq H^\can_1$, we only need to establish that there is a concentrating
sequence of projections for which its $H-$function equals the Gibbs-von Neumann entropy. Hence,
the proof of Theorem \ref{thm: equivalence} follows from the following lemma:
\begin{lemma}
If a sequence of states $(\om^N)$ satisfies conditions i) and ii) of Theorem~\ref{thm:
equivalence}, then there exists a sequence of projections $(P^N)$ exponentially concentrating at
$x$ and satisfying
\begin{equation}\label{eq: equiv-lem}
  \lim_{N\uparrow +\infty} \frac{1}{N} (\log\Tr^N(P^N) - \caH(\om^N)) = 0
\end{equation}
\end{lemma}
\begin{proof}
There exists a sequence $\de_N \downarrow 0$ such that when substituted for $\delta$,
\eqref{eq: AEP} is still satisfied.  Take such a sequence and define
$P^N = \int_{-\de_N}^{\de_N} \id\tilde Q^N(z)$. By construction,
\begin{equation}\label{eq: op-bounds}
  e^{N(h_N - \de_N)}P^N \leq (\si^N)^{-1} P^N \leq e^{N(h_N + \de_N)}P^N
\end{equation}
for any $N = 1,2,\ldots$, with the shorthand
$h_N = \frac{1}{N} \caH(\om^N)$. That yields the inequalities
\begin{align}
  \Tr^N(P^N) &= \om^N((\si^N)^{-1}P^N) \leq e^{N(h_N + \de_N)}\om^N(P^N)
\\\intertext{and}\label{eq: lower bound}
  \Tr^N(P^N) &\geq e^{N(h_N - \de_N)}\om^N(P^N)
\end{align}
Using that $\lim_{N \uparrow +\infty} \frac{1}{N} \log\om^N(P^N) = 0$ proves \eqref{eq:
equiv-lem}.

To see that $(P^N)$ is exponentially concentrating at $x$, observe that for all $Y^N \geq 0$,
\begin{equation}
\begin{split}
  \om^N(Y^N) &= \Tr^N((\si^N)^\frac{1}{2} Y^N (\si^N)^\frac{1}{2})
\\
  &\geq \Tr^N(P^N (\si^N)^\frac{1}{2} Y^N (\si^N)^\frac{1}{2} P^N)
\\
  &= \Tr^N((Y^N)^\frac{1}{2} P^N \si^N (Y^N)^\frac{1}{2})
\\
  &\geq e^{N(h_N - \de_N)} \Tr^N(P^N)\, \tr^N(Y^N \rel P^N)
\\
  &\geq e^{-2 N \de_N} \om^N(P^N)\,\tr^N(Y^N \rel P^N)
\end{split}
\end{equation}
where we used inequalities~\eqref{eq: op-bounds}-\eqref{eq: lower bound}. By the exponential
concentration property of $(\om^N)$, inequality \eqref{excp}, for all
$k \in K$, $\ep > 0$, and $N > N_k(\ep)$
\begin{equation}\label{eq: proof-exp. concentration}
  \int_{\bbR \setminus (x_k - \ep, x_k + \ep)} \tr^N(\id Q^N_k(z) \rel P^N)
  \leq e^{-(C_k(\ep) - 2\de_N)N} (\om^N(P^N))^{-1}
\end{equation}
Choose $N'_k(\ep)$ such that $\de_N \leq \frac{C_k(\ep)}{8}$ and
$\frac{1}{N} \log\om^N(P^N) \geq -\frac{C_k(\ep)}{4}$ for all $N >
N'_k(\ep)$. Then \eqref{eq: proof-exp. concentration} $\leq
\exp[-\frac{C_k(\ep) N}{2}]$ for all $N > \max\{N_k(\ep),
N'_k(\ep)\}$.
\end{proof}

\section{$H-$theorem from macroscopic autonomy}\label{sec: H-theorems}

When speaking about an $H-$theorem or about the monotonicity of entropy one often refers, and even
more so for a quantum set-up, to the fact that the relative entropy verifies the contraction
inequality
\begin{equation}\label{relmon}
  \caH(\om^N \tau^N \rel \rho^N \tau^N) \leq \caH(\om^N \rel \rho^N)
\end{equation}
for all states $\omega^N, \rho^N$ on $\mathscr{H}^N$ and for all completely positive maps $\tau^N$
on $\mathscr{B}(\mathscr{H}^N)$. That is true classically, quantum mechanically and for all small
or large $N$. When the reference state $\rho^N$ is invariant under
$\tau^N$, \eqref{relmon} yields the contractivity of the relative
entropy with respect to $\rho^N$. However tempting, such inequalities should not be confused with
second law or with
$H-$theorems; note in particular that $\caH(\om^N)$ defined
in~\eqref{eq: G.-von N.} is constant whenever $\tau^N$ is an automorphism: $\caH(\om^N \tau^N) =
\caH(\om^N)$.

In contrast, an $H-$theorem refers to the (usually strict) monotonicity of a quantity on the
macroscopic trajectories as obtained from a microscopically defined dynamics. Such a quantity is
often directly related to the fluctuations in a large system and its extremal value corresponds to
the equilibrium or, more generally, to a stationary state.

In the previous Section we have obtained how to represent a macroscopic state and constructed a
candidate $H-$function. Imagine now a time-evolution for the macroscopic values, always referring
to the same set of (possibly noncommuting macroscopic) observables $X^N_k$. To prove an
$H-$theorem, we need basically two assumptions: macroscopic autonomy and the semigroup property, or
that there is a first order autonomous equation for the macroscopic values. A classical version of
this study and more details can be found in \cite{dmn}.

\subsection{Microcanonical set-up}
Assume a family of automorphisms $\tau^N_{t,s}$ is given as acting on the observables from
$\mathscr{B}(\mathscr{H}^N)$ and satisfying
\begin{equation}
  \tau_{t,s}^N=\tau_{t,u}^N\,\tau^N_{u,s}
  \qquad t \geq u \geq s
\end{equation}
It follows that the trace $\Tr^N$ is invariant for $\tau_{t,s}^N$.\\
Recall that $\Om \subset \bbR^K$ is the set of all admissible macroscopic configurations,
$H^{\mc}(x) \geq 0$. On this space we want to study the emergent macroscopic dynamics.

\vspace{3mm}
\noindent{\bf Autonomy condition}\\
There are maps $(\phi_{t,s})_{t \geq s \geq 0}$  on $\Om$ and there is a microcanonical macrostate
$(P^N)$, $P^N=P^N (x)$ for each $x \in \Om$,  such that for all $G \in \caF$ and
$t \geq s \geq 0$,
\begin{equation}\label{eq: strong autonomy mc}
  \lim_{N\uparrow +\infty} \tr^N(\tau_{t,s}^N G(X^N) \rel P^N) = G(\phi_{t,s} x)
\end{equation}
\noindent{\bf Semigroup property}\\
The maps are required to satisfy the semigroup condition,
\begin{equation}\label{eq: markov property}
  \phi_{t,u}\, \phi_{u,s} = \phi_{t,s}
\end{equation}\label{eq: semigroup property}
for all $t \geq u \geq s \geq 0$.

\begin{theorem}\label{thm: H-theorem mc}
Assume that the autonomy condition \eqref{eq: strong autonomy mc}
and the semigroup condition \eqref{eq: markov property} are both
satisfied. Then, for every $x \in \Om$, $H^{\mc}(x_t)$ is
nondecreasing in $t \geq 0$ with $x_t = \phi_{t,0}x$.
\end{theorem}

\begin{proof}
Given $x \in \Om$, fix a microcanonical macrostate $P^N \tomc x$ and $t \geq s \geq 0$. Using that
$(\tau^N_{t,s})^{-1}$ is an automorphism and
$\Tr^N((\tau^N_{t,s})^{-1} \cdot) = \Tr^N(\cdot)$, the identity
\[
  \tr^N(\tau^N_{t,s} G(X^N) \rel P^N) =
  \frac{\Tr^N(G(X^N) (\tau_{t,s}^N)^{-1} P^N)}{\Tr^N((\tau_{t,s}^N)^{-1} P^N)}
  = \tr^N(G(X^N) \rel (\tau_{t,s}^N)^{-1} P^N)
\]
yields $(\tau^N_{t,s})^{-1} P^N \tomc \phi_{t,s} x$ due to autonomy condition~\eqref{eq: strong
autonomy mc}. Hence,
\[
  H^{\mc}(\phi_{t,s} x) \geq \limsup_{N\uparrow +\infty}
  \frac{1}{N} \log\Tr^N((\tau^N_{t,s})^{-1} P^N) = H^{\mc}(x)
\]
In particular, one has that $x_s = \phi_{s,0} x \in \Om$. The statement then follows by the
semigroup property~\eqref{eq: strong autonomy mc}:
\[
  H^{\mc}(x_t) = H^{\mc}(\phi_{t,0} x) = H^{\mc}(\phi_{t,s} x_s) \geq H^{\mc}(x_s)
\]
\end{proof}

It is important to realize that a macroscopic dynamics, even autonomous in the sense of \eqref{eq:
strong autonomy mc}, need not satisfy the semigroup property \eqref{eq: semigroup property}. In
that case one actually does not expect the $H-$function to be monotone; see \cite{tim} and below
for an example. As obvious from the proof, without that semigroup property of
$(\phi_{t,s})$, \eqref{eq: strong autonomy mc} only implies
$H(x_t) \geq H(x)$, $t \geq 0$. Or, in a bit more generality, it
implies that for all $s \geq 0$ and $x \in \Om$ the macrotrajectory $(x_t)_{t \geq s}$, $x_t =
\phi_{t,s}(x)$ satisfies $H(x_t) \geq H(x_s)$ for all $t \geq s$.

Remark that while the set of projections is invariant under the automorphisms $(\tau^N_{t,s})$,
this is not true any longer for more general microscopic dynamics defined as completely positive
maps, and describing possibly an open dynamical system interacting with its environment. In the
latter case the proof of Theorem~\ref{thm: H-theorem mc} does not go through and one has to allow
for macrostates described via more general states, as in Section~\ref{cans}. The revision of the
argument for the
$H-$theorem within the canonical set-up is done in the next
section.

\subsection{Canonical set-up}

We have completely positive maps $(\tau^N_{t,s})_{t \geq s \geq 0}$ on
$\mathscr{B}(\mathscr{H}^N)$ satisfying
\begin{equation}
  \tau_{t,s}^N=\tau_{t,u}^N\,\tau^N_{u,s}
  \qquad t \geq u \geq s \geq 0
\end{equation}
and leaving invariant the state $\rho^N$; they represent the microscopic dynamics. The macroscopic
dynamics is again given by maps $\phi_{t,s}$.\\

As a variant of autonomy condition \eqref{eq: strong autonomy mc},
we assume that the maps $\phi_{t,s}$ are reproduced along the
time-evolution in the mean. Namely, see definition \eqref{addi},
for every $x \in \Om_1(\rho) = \{x;\,H^\can_1(x \rel \rho) <
\infty\}$ we ask that a canonical macrostate $\om^N \tomean x$
exists such that, for all $t \geq s \geq 0$,
\begin{equation}\label{def: strong autonomy can}
  \phi_{t,s} x =  \lim_{N\uparrow +\infty} \omega^N(\tau^N_{t,s} X^N)
\end{equation}
At the same time, we still assume the semigroup condition \eqref{eq: markov property}.

\begin{theorem}\label{thm: H=theorem can}
Under conditions \eqref{def: strong autonomy can} and \eqref{eq: markov property}, the function
$H^{\can}_1(\phi_{t,0} x \rel \rho)$ is nonincreasing in $t \geq 0$ for all $x \in \Om_1(\rho)$.
\end{theorem}
\begin{proof}
If $\om^N \tomean x$ is a canonical macrostate at $x$ then, by the monotonicity of the relative
entropy,
\[
  H^{\can}_1(x \rel \rho) = \liminf_{N \uparrow +\infty} \frac{1}{N} \caH ( \omega^N \rel \rho^N)
  \geq \liminf_{N \uparrow +\infty} \frac{1}{N} \caH ( \omega^N \tau^N_{t,s} \rel \rho^N)
\]
On the other hand, by \eqref{def: strong autonomy can}, the sequence $(\omega^N \tau^N_{t,s})$ is
concentrating in the mean at $\phi_{t,s} (x)$, yielding
\begin{equation*}
  H^{\can}_1(x \rel \rho) \geq  H^{\can}_1(\phi_{t,s} x \rel \rho)
\end{equation*}
Using~\eqref{eq: markov property}, the proof is now finished as in Theorem \ref{thm: H-theorem
mc}.
\end{proof}

\subsection{Example: the quantum Kac model}

A popular toy model to illustrate and to discuss essential features of relaxation to equilibrium
has been introduced by Mark Kac, \cite{Kac}. Here we review an extension that can be called a
quantum Kac model, we described it extensively in \cite{tim}, to learn only later that essentially
the same model was considered by Max Dresden and Frank Feiock in \cite{dresden}. However, there is
an interesting difference in interpretation to which we return at the end of the section.

At each site of a ring with $N$ sites there is a quantum bit
$\psi_i\in \mathbb{C}^2$ and a classical binary variable $\xi_i =
\pm 1$ (which we also consider to be embedded in $\mathbb{C}^2$).
The microstates are thus represented as vectors $(\psi;\xi)=
(\psi_1,\ldots,\psi_N;\xi_1,\ldots,\xi_N)$, being elements of the Hilbert space $\mathscr{H}^N =
\bbC^{2N} \otimes \bbC^{2N}$. The time is discrete and at each step two operations are performed:
a right shift, denoted below by $S^N$ and a local scattering or update $V^N$. The unitary dynamics
is given as
\begin{equation}
U ^N = S^N V^N   \qquad  U_t^N = (U^N)^t   \textrm{ for } t \in \bbN
\end{equation}
with the shift
\begin{equation}
S^N(\psi;\xi)=(\psi_N,\psi_1,\ldots,\psi_{N-1};\xi)
\end{equation}
and the scattering
\begin{equation}
V^N(\psi;\xi) =  (\frac{1-\xi_1}{2}V_1\psi_1 +
\frac{1+\xi_1}{2}\psi_1,\ldots,\frac{1-\xi_N}{2}V_N\psi_N +
\frac{1+\xi_N}{2}\psi_N;\xi)
\end{equation}
extended to an operator on $\mathscr{H}^N$ by linearity. Here, $V$ is a unitary $2\times2$ matrix
and $V_i$ its copy at site $i = 1,\ldots,N$.

We consider the family of macroscopic observables
\[
X_0^N = \frac 1{N}\sum_{i=1}^N \xi_i,\qquad  X_\alpha^N =
\frac 1{N}\sum_{i=1}^N \sigma_i^\alpha \quad \alpha=1,2,3
\]
where $\sigma_i^1, \si_i^2, \si_i^3$ are the Pauli matrices acting at site $i$ and embedded to
operators on $\mathscr{H}^N$. We fix macroscopic values
$x=(\mu,m_1,m_2,m_3) \in [-1,+1]^4$ and we construct a
microcanonical macrostate $(P^N)$ in $x$ in the following way.\\
Let $\delta_N$ be a positive sequence in $\bbR$ such that $\de_N
\downarrow 0$ and $N^{1/2}\de_N \uparrow +\infty$ as $N\uparrow
+\infty$. For $\mu \in [-1,1]$, let $Q_{0}^{N}(\mu)$ be the spectral projection associated to
$X_0^N$, on the interval
$[\mu-\de_N,\mu+\de_N]$. For $\vec{m}=(m_1,m_2,m_3) \in [-1,1]^3$,
we already constructed a microcanonical macrostate $Q^{N}(\vec m)$ in Section 2.1.4. Obviously,
$Q_{0}^N(\mu)$ and $Q^N(\vec m)$ commute and the product $P^N = Q_{0}^N(\mu)\, Q^N(\vec m)$ is a
projection.
It is easy to check that $P^{N}$ is a microcanonical macrostate at $x=(\mu,\vec{m})$.\\
The construction of the canonical macrostate is standard along the lines of Section~\ref{cmex}.
The corresponding $H-$functions are manifestly equal:
\begin{equation}H^{\mc}(x) = H^\can_1(x) =
\eta\bigl(\frac{1+m}{2}\bigr)+\eta\bigl(\frac{1-m}{2}\bigr)+
\eta\bigl(\frac{1+\mu}{2}\bigr)+\eta\bigl(\frac{1-\mu}{2}\bigr)
\end{equation}
with $\eta(x) = -x \log x$ for $x \in (0,1]$ and $\eta(0) = 0$, otherwise
$\eta(x) = -\infty$.

We now come to the conditions of Theorem \ref{thm: H-theorem mc}. The construction of the
macroscopic dynamics and the proof of its autonomy was essentially done in
\cite{tim}. The macroscopic equation $\xi_t=\xi$ is obvious and the equation for
$\vec{m}_t$ can be written, associating $\vec m_t$ with the reduced $2\times 2$ density matrix
$\nu_t = (1 + \vec{m}_t\cdot \vec{\sigma})/2$, in the form
$\nu_t = \Lambda_\mu^t \nu$, $t=0,1,\ldots$, where
$\La_\mu^t = (\La_\mu)^t$ and
\begin{equation}
\Lambda_{\mu}(\nu) =  \frac{1-\mu}{2} V \nu V^* +
\frac{1+\mu}{2}\nu
\end{equation}
The semigroup condition \eqref{eq: markov property} is then also automatically checked.\\
In order to understand better the necessity of the semigroup property for an $H-$theorem to be
true, compare the above with another choice of macroscopic variables. Assume we had started out
with
\[
X_0^N = \frac 1{N}\sum_{i=1}^N \xi_i,\quad  X_1^N =
\frac 1{N}\sum_{i=1}^N \sigma_i^1
\]
as the only macroscopic variables, as was done in \cite{dresden}. A microcanonical macrostate can
again be easily constructed by setting $Q_{0}^{N}(\mu)$ the spectral projection associated to
$X_0^N$ on the interval $[\mu-\de_N,\mu +\de_N]$ and
$Q_{1}^{N}(\vec m)$ the spectral projection for $X_1^N$ on
$[\mu-\de_N,\mu +\de_N]$, and finally $P^{N} = Q_{0}^{N}(\mu)\,
Q_{1}^{N}(\vec m)$ as before. The sequence $(P^N)$ defines a microcanonical macrostate at
$(\mu,\vec m)$ and the autonomy condition \eqref{eq: strong autonomy mc} is satisfied.  However,
the macroscopic evolution does not satisfy the semigroup property
\eqref{eq: markov property} and, in agreement with that, the
corresponding $H-$functions are not monotonous in time (see
\cite{tim}).

\vspace{5mm}
\noindent{\bf Acknowledgment}\\
We thank Andr\'e Verbeure for useful discussions and for his ongoing interest in this line of
research.  K.N. acknowledges the support from the project AVOZ10100520 in the Academy  of Sciences
of the Czech Republic.

\bibliographystyle{plain}

\end{document}